\documentclass[12pt]{article}

\usepackage{amssymb}
\usepackage{amsmath}
\usepackage{amsfonts}
\usepackage{amsbsy}

\begin{document}

\begin{titlepage}

\begin{center}

\vspace*{-10ex} \hspace*{\fill} KAIST-TH/2005-19

\vskip 1.5cm

\Huge{From the Spectrum to Inflation:\\
A Second Order Inverse Formula for\\
the General Slow-Roll Spectrum}

\vskip 1cm

\large{
Minu Joy\footnote{minujoy@muon.kaist.ac.kr}
and
Ewan D. Stewart
\\ \vspace{0.5cm}
{Department of Physics, KAIST, Daejeon, Republic of Korea}
}

\vskip 0.5cm

\today

\vskip 1.2cm

\end{center}

\begin{abstract}
We invert the second order, single field, general slow-roll formula for the power spectrum, to obtain a second order formula for inflationary parameters in terms of the primordial power spectrum.
\end{abstract}

\end{titlepage}

\setcounter{page}{0}

\newpage

\setcounter{page}{1}

\section{Introduction}

The primordial curvature perturbations \cite{ChibisovMukhanov} are manifested today by
anisotropies in the cosmic microwave background (CMB) and inhomogeneities in the large scale
structure of the observable universe, and are characterised in terms of the power spectrum.
Observations \cite{obs} of the CMB and large scale structure allow the reconstruction of
inflationary parameters \cite{recon1,recon2,recon3}. For a proper analysis of the upcoming high
precision observations, such as Planck \cite{planck}, we need accurate analytical studies.
Usually, one chooses a particular theory, assesses the spectrum it predicts and attempts a
comparison between its predictions and the observed universe. However, to interpret observations,
it is more useful to have an inverse formula that determines inflationary parameters in terms of
the primordial spectrum, which is in turn determined by observational quantities.

The standard slow-roll approximation used for inflationary scenarios makes some strong assumptions
about the properties of inflation, which have not yet been confirmed observationally. Hence, a
more general slow-roll approximation has been put forward \cite{gsr} which lifts the
extra, unjustified, assumptions of the standard slow-roll approximation. The advantages of the
general slow-roll approximation, compared with the standard approximation, are clearly discussed
in Ref.~\cite{ewanrecon}.

Recently we proposed a general inverse formula \cite{inverse} for extracting inflationary parameters from the power spectrum of cosmological perturbations.
There, we inverted the leading order, single field, general slow-roll formula for the power spectrum to obtain a formula for inflationary parameters in terms of the primordial power spectrum.
Here we extend our inverse formalism to the second order.

\section{Second order general slow-roll power spectrum}

We follow the formalism of our first inverse paper \cite{inverse}. For single field inflationary
models, it is convenient to express inflationary quantities in terms of \cite{gsr,jo1}
\begin{equation}\label{f}
f = \frac{2\pi a \xi \dot\phi}{H}
\end{equation}
where $\xi = - \int \frac{dt}{a} = \frac{1}{aH} \left( 1 - \frac{\dot{H}}{H^2} + \ldots \right)$
is minus the conformal time. We think of $f$ as a function of $\ln\xi$ so that $f' \equiv
df/d\ln\xi$.

Under the general slow-roll approximation, the spectrum for a single field inflation model can be
expressed up to second order terms as \cite{secondorder}
\begin{eqnarray}\label{SOP}
\ln\mathcal{P}(\ln k) & = &
\int_0^\infty \frac{d\xi}{\xi} \left[ - k\xi \, W'(k\xi) \right]
\left[ \ln\frac{1}{f(\ln\xi)^2} + \frac{2}{3} \frac{f'(\ln\xi)}{f(\ln\xi)} \right]
+ \frac{\pi^2}{2} \left[ \int_0^\infty \frac{d\xi}{\xi} m(k\xi) \frac{f'(\ln\xi)}{f(\ln\xi)} \right]^2
\nonumber \\ && \mbox{}
- 2\pi \int_0^\infty \frac{d\xi}{\xi} m(k\xi) \frac{f'(\ln\xi)}{f(\ln\xi)}
\int_\xi^\infty \frac{d\zeta}{\zeta} \frac{1}{k\zeta} \frac{f'(\ln\zeta)}{f(\ln\zeta)}
\end{eqnarray}
where
\begin{equation}\label{W}
W(x) = \frac{3\sin(2x)}{2x^3} - \frac{3\cos(2x)}{x^2} - \frac{3\sin(2x)}{2x} - 1
\end{equation}
and
\begin{equation}
m(x) = \frac{2}{\pi} \left[ \frac{1}{x} - \frac{\cos(2x)}{x} - \sin(2x) \right]
\end{equation}
$W(x)$ and $m(x)$ have the asymptotic behaviors
\begin{equation}
\lim_{x \rightarrow 0} W(x) = \frac{2}{5} x^2 + \mathcal{O}(x^4)
\end{equation}
and
\begin{equation}
\lim_{x \rightarrow 0} m(x) = \frac{4}{3\pi} x^3 + \mathcal{O}(x^5)
\end{equation}
and window properties
\begin{equation}
\int_0^\infty \frac{dx}{x} \left[ - x \, W'(x) \right] = 1
\end{equation}
\begin{equation}\label{mwin}
\int_0^\infty \frac{dx}{x} \, m(x) = 1
\end{equation}
and
\begin{equation}
\int_0^\infty \frac{dx}{x} \, \frac{1}{x} \, m(x) = \frac{2}{\pi}
\end{equation}

\subsection{Inverse formula}

Our fundamental inverse identity is
\begin{equation}\label{id}
\int_0^\infty \frac{dk}{k} \, m(k\zeta) \, W(k\xi)
= \frac{1}{2} \frac{\zeta^3}{\xi^3} \left[ \mathrm{sgn}(\xi+\zeta) + \mathrm{sgn}(\xi-\zeta) \right]
+ \frac{1}{2} \left[ \mathrm{sgn}(\xi+\zeta) - \mathrm{sgn}(\xi-\zeta) \right]
- \mathrm{sgn}(\zeta)
\end{equation}
where $\mathrm{sgn}(x)=-1$ for $x<0$ and $\mathrm{sgn}(x)=1$ for $x>0$.
Taking the derivative with respect to $\xi$ we get
\begin{equation}\label{id2}
\int_0^\infty \frac{dk}{k} \, m(k\zeta) \left[ - k\xi \, W'(k\xi) \right]
= \frac{3}{2} \frac{\zeta^3}{\xi^3} \left[ \mathrm{sgn}(\xi+\zeta) + \mathrm{sgn}(\xi-\zeta) \right]
\end{equation}
which we can use to invert the first order general slow-roll formula \cite{gsr},
\begin{equation}
\ln\mathcal{P} = \int_0^\infty \frac{d\xi}{\xi} \left[ - k\xi \, W'(k\xi) \right] \left(
\ln\frac{1}{f^2} + \frac{2}{3} \frac{f'}{f} \right)
\end{equation}
to get \cite{inverse}
\begin{equation}
\ln\frac{1}{f^2} = \int_0^\infty \frac{dk}{k} \, m(k\xi) \ln\mathcal{P}
\end{equation}
and hence
\begin{equation}
2 \frac{f'}{f} = \int_0^\infty \frac{dk}{k} \, m(k\xi) \, \frac{\mathcal{P}'}{\mathcal{P}}
\end{equation}
Substituting into Eq.~(\ref{SOP}) gives
\begin{eqnarray}
\lefteqn{
\int_0^\infty \frac{d\xi}{\xi} \left[ - k\xi \, W'(k\xi) \right] \left[
\ln\frac{1}{f(\ln\xi)^2} + \frac{2}{3} \frac{f'(\ln\xi)}{f(\ln\xi)} \right]
} \nonumber \\
& = & \ln\mathcal{P}(\ln k)
- \frac{\pi^2}{8} \left[ \int_0^\infty \frac{d\xi}{\xi} m(k\xi) \int_0^\infty
\frac{dl}{l} \, m(l\xi) \, \frac{\mathcal{P}'(\ln l)}{\mathcal{P}(\ln l)} \right]^2
\nonumber \\
&& \mbox{} + \frac{\pi}{2} \int_0^\infty \frac{d\xi}{\xi} m(k\xi) \int_0^\infty \frac{dl}{l} \,
m(l\xi) \, \frac{\mathcal{P}'(\ln l)}{\mathcal{P}(\ln l)} \int_\xi^\infty \frac{d\zeta}{\zeta}
\frac{1}{k\zeta} \int_0^\infty \frac{dq}{q} \, m(q\zeta) \, \frac{\mathcal{P}'(\ln
q)}{\mathcal{P}(\ln q)}
\end{eqnarray}
and applying Eq.~(\ref{id2}) gives
\begin{eqnarray}
\lefteqn{
\ln\frac{1}{f(\ln\xi)^2} =
\int_0^\infty \frac{dk}{k} \, m(k\xi) \ln\mathcal{P}(\ln k)
- \frac{\pi^2}{8} \int_0^\infty \frac{dk}{k} \, m(k\xi)
\left[ \int_0^\infty \frac{d\zeta}{\zeta} m(k\zeta) \int_0^\infty \frac{dl}{l} \, m(l\zeta) \, \frac{\mathcal{P}'(\ln l)}{\mathcal{P}(\ln l)} \right]^2
} \nonumber \\ && \mbox{}
+ \frac{\pi}{2} \int_0^\infty \frac{dk}{k} \, m(k\xi)
\int_0^\infty \frac{d\zeta}{\zeta} m(k\zeta)
\int_0^\infty \frac{dl}{l} \, m(l\zeta) \, \frac{\mathcal{P}'(\ln l)}{\mathcal{P}(\ln l)}
\int_\zeta^\infty \frac{d\chi}{\chi} \frac{1}{k\chi}
\int_0^\infty \frac{dq}{q} \, m(q\chi) \, \frac{\mathcal{P}'(\ln q)}{\mathcal{P}(\ln q)}
\nonumber \\
\end{eqnarray}
Reordering we get
\begin{eqnarray}\label{soinv1}
\lefteqn{
\ln\frac{1}{f(\ln\xi)^2} =
\int_0^\infty \frac{dk}{k} \, m(k\xi) \ln\mathcal{P}(\ln k)
- \frac{\pi^2}{8} \int_0^\infty \frac{dk}{k} \, m(k\xi)
\left[ \int_0^\infty \frac{dl}{l} \frac{\mathcal{P}'(\ln l)}{\mathcal{P}(\ln l)}
\int_0^\infty \frac{d\zeta}{\zeta} \, m(k\zeta) \, m(l\zeta) \right]^2
} \nonumber \\ && \mbox{}
+ \frac{\pi}{2} \int_0^\infty \frac{dl}{l} \frac{\mathcal{P}'(\ln l)}{\mathcal{P}(\ln l)}
\int_0^\infty \frac{dq}{q} \frac{\mathcal{P}'(\ln q)}{\mathcal{P}(\ln q)}
\int_0^\infty \frac{d\zeta}{\zeta} \, m(l\zeta)
\int_0^\infty \frac{dk}{k^2} \, m(k\xi) \, m(k\zeta)
\int_\zeta^\infty \frac{d\chi}{\chi^2} \, m(q\chi)
\nonumber \\
\end{eqnarray}
Now, we have the identity
\begin{equation}\label{2}
\int_0^\infty \frac{d\zeta}{\zeta} \, m(k\zeta) \, m(l\zeta) = \frac{2}{\pi^2}
\ln\left|\frac{k+l}{k-l}\right|
\end{equation}
and using the identities
\begin{equation}
\int_\zeta^\infty \frac{d\chi}{\chi^2} \, m(q\chi) = \frac{2}{\pi}\frac{\sin^2(q\zeta)}
{q\zeta^2}
\end{equation}
and
\begin{equation}
\int_0^\infty  \frac{dk}{k^2} \, m(k\xi) \, m(k\zeta)
= \frac{2}{3\pi\xi\zeta} \left[ 2\xi^3 \, \mathrm{sgn}(\xi)
+ 2\zeta^3 \, \mathrm{sgn}(\zeta)
- (\xi^3 - \zeta^3) \, \mathrm{sgn}(\xi-\zeta)
- (\xi^3 + \zeta^3) \, \mathrm{sgn}(\xi+\zeta) \right]
\end{equation}
we get
\begin{equation}\label{3}
\int_0^\infty \frac{d\zeta}{\zeta} \, m(l\zeta) \, \int_0^\infty \frac{dk}{k^2} \,
m(k\xi) \, m(k\zeta) \, \int_\zeta^\infty \frac{d\chi}{\chi^2} \, m(q\chi)
= \frac{2}{\pi} \, M(l\xi,q\xi)
\end{equation}
where
\begin{eqnarray}
M(x,y) & = & \frac{2}{\pi^2 x y} \left[ g(x) + g(y) - \frac{1}{2} \, g(x-y) - \frac{1}{2} \,
g(x+y) \right]
\end{eqnarray}
with
\begin{equation}
g(x) = x \, \mathrm{Si}(2x) + \frac{\cos(2x)}{2} - \frac{1}{2}
\end{equation}
where
\begin{equation}
\mathrm{Si}(x) \equiv \int_0^x \frac{\sin t}{t} \, dt
\end{equation}
$M(x,y)$ has the window property
\begin{equation}\label{Mwin}
\int_0^\infty \frac{dx}{x} \int_0^\infty \frac{dy}{y} \, M(x,y) = 1
\end{equation}
and the asymptotic behaviour
\begin{equation}
\lim_{x,y \rightarrow 0} M(x,y) = \frac{4 x y}{3\pi^2} \left[ 1 + \mathcal{O}\left(x^2+y^2
\right)\right]
\end{equation}
Substituting Eqs.~(\ref{2}) and~(\ref{3}) into Eq.~(\ref{soinv1}), we get the simplified form for the second order inverse formula
\begin{eqnarray}\label{soinv}
\ln\frac{1}{f(\ln\xi)^2} & = &
\int_0^\infty \frac{dk}{k} \, m(k\xi) \ln\mathcal{P}(\ln k)
\nonumber \\ && \mbox{}
- \frac{1}{2\pi^2} \int_0^\infty \frac{dk}{k} \, m(k\xi)
\left[ \int_0^\infty \frac{dl}{l} \ln\left|\frac{k+l}{k-l}\right| \frac{\mathcal{P}' (\ln l)}{\mathcal{P}(\ln l)} \right]^2
\nonumber \\ && \mbox{}
+ \int_0^\infty \frac{dl}{l} \int_0^\infty \frac{dq}{q} \, M(l\xi,q\xi) \,  \frac{\mathcal{P}'(\ln l)}{\mathcal{P}(\ln l)} \frac{\mathcal{P}'(\ln q)}{\mathcal{P}(\ln q)}
\end{eqnarray}

\section{Examples}

\subsection{Standard slow-roll approximation}

In the context of standard slow-roll, the power spectrum has the form
\begin{equation}
\ln\mathcal{P} = \ln\mathcal{P}_\diamond + (n_\diamond - 1) \ln \left( \frac{k}{k_\diamond} \right) + \frac{1}{2} n'_\diamond \ln^2 \left(\frac{k}{k_\diamond}\right)  + \cdots
\end{equation}
where $n$ is the spectral index and $k_\diamond$ is some reference wavenumber.
Applying our inverse formula Eq.~(\ref{soinv}), using the window properties Eqs.~(\ref{mwin}) and~(\ref{Mwin}), and the results
\begin{equation}
\int_0^\infty \frac{dx}{x} \, m(x) \, \ln(x) = \alpha
\end{equation}
\begin{equation}
\int_0^\infty \frac{dx}{x} \, m(x) \, \ln^2(x) = \alpha^2 + \frac{\pi^2}{12}
\end{equation}
\begin{equation}
\frac{1}{2\pi^2} \int_0^\infty \frac{dx}{x} \, m(x) \left[ \int_0^\infty \frac{dy}{y} \ln\left|\frac{x+y}{x-y}\right| \right]^2 = \frac{\pi^2}{8}
\end{equation}
where $\alpha = 2 - \ln 2 - \gamma \simeq 0.7296$, gives
\begin{equation}\label{ssrinv}
\ln\frac{1}{f^2} = \ln\mathcal{P}_\diamond + \alpha_\diamond (n_\diamond-1)
+ \frac{1}{2} \left( \alpha_\diamond^2 + \frac{\pi^2}{12} \right) n'_\diamond
+ \left( 1 - \frac{\pi^2}{8} \right) (n_\diamond-1)^2 + \cdots
\end{equation}
where $\alpha_\diamond = \alpha - \ln(k_\diamond \xi)$.
Eq.~(\ref{ssrinv}) reproduces the standard slow-roll inverse, which is trivially obtained from the standard slow-roll formula \cite{jo1}
\begin{equation}
\ln\mathcal{P} =
\ln\frac{1}{f_\star^2} - 2 \alpha_\star \frac{f_\star'}{f_\star}
- \left( \alpha^2_\star - \frac{\pi^2}{12} \right) \frac{f_\star''}{f_\star}
+ \left(\alpha^2_\star - 4 + \frac{5\pi^2}{12} \right) \left(\frac{f_\star'}{f_\star}\right)^2 + \cdots
\end{equation}
where $\alpha_\star = \alpha - \ln(k \xi_\star)$ and $\xi_\star$ is an arbitrary evaluation point usually taken to be around horizon crossing.

\subsection{Power law}

Consider the simple case where the spectrum has the power law form
\begin{equation}\label{powerP}
\ln \mathcal{P} = \ln \mathcal{P}_0 - A_1 k^\nu - A_2 k^{2\nu}
\end{equation}
with $\nu > 0$.
Substituting this power spectrum into our inverse formula Eq.~(\ref{soinv}), and using the results
\begin{equation}\label{C}
\int_0^\infty \frac{dx}{x} \, m(x) \, x^\nu
= \frac{2^{1-\nu}}{\pi} \frac{\Gamma(2+\nu)}{\nu(1-\nu)} \sin \left( \frac{\pi\nu}{2} \right)
\equiv C(\nu)
\end{equation}
\begin{equation}
\frac{\nu^2}{2\pi^2} \int_0^\infty \frac{dx}{x} \, m(x) \left[\int_0^\infty \frac{dy}{y} \ln\left|\frac{x+y}{x-y}\right| y^\nu \right]^2
= \frac{1}{2} \tan^2\left(\frac{\pi\nu}{2}\right) C(2\nu)
\end{equation}
and
\begin{equation}
\int_0^\infty \frac{dx}{x} \int_0^\infty \frac{dy}{y} \, M(x,y) \, x^\nu y^\nu
= \frac{C^2(\nu)}{(1-2\nu)(1+\nu)^2}
\end{equation}
we obtain
\begin{eqnarray}\label{powerfP}
\ln \frac{1}{f^2} & = & \ln\mathcal{P}_0 - A_1 \, C(\nu) \, \xi^{-\nu}
\nonumber \\ && \mbox{}
- \left[ A_2 \, C(2\nu) + \frac{1}{2} A_1^2 \tan^2\left(\frac{\pi\nu}{2}\right) C(2\nu) - \frac{A_1^2 \nu^2}{(1-2\nu)(1+\nu)^2} \, C^2(\nu) \right] \xi^{-2\nu}
\end{eqnarray}

To understand this result better, let us see the forward calculation.
Consider an inflationary scenario with
\begin{equation}\label{powerf}
\ln \frac{1}{f^2}  = \ln  \frac{1}{f_\infty^2}  - B_1 \xi^{-\nu} - B_2 \xi^{-2\nu}
\end{equation}
Calculating the power spectrum up to second order terms using Eq.~(\ref{SOP}) we obtain
\begin{equation}\label{powerPf}
\ln\mathcal{P}  =  \ln\frac{1}{f_\infty^2} - \frac{B_1}{C(\nu)} k^\nu - \left[ \frac{B_2}{C(2\nu)} - \frac{1}{2} \tan^2\left(\frac{\pi\nu}{2}\right) \, \frac{B_1^2}{C^2(\nu)} + \frac{\nu^2}{(1-2\nu)(1+\nu)^2} \, \frac{B_1^2}{C(2\nu)} \right] k^{2\nu}
\end{equation}
The term proportional to $A_1$ in Eq.~(\ref{powerfP}), \mbox{i.e.} the first order result, diverges at $\nu=1$.
This is to be expected as the term proportional to $B_1$ in Eq.~(\ref{powerPf}), \mbox{i.e.} the first order general slow-roll result, is degenerate at $\nu=1$ and so one can not expect an inverse to exist beyond that point.
The second order terms in Eq.~(\ref{powerfP}) diverge at $\nu=1/2$.
This is to be expected as the term proportional to $B_2$ in Eq.~(\ref{powerPf}) is degenerate at $\nu=1/2$ and so one can not expect a second order inverse to exist beyond that point.

\subsection{Linear potential with a sharp slope change}

Consider a linear potential with slope changing from $-A$ to $-A - \Delta A$ at $\phi = \phi_0$ \cite{Star,secondorder,scaledepn}
\begin{equation}
V(\phi) = V_0 \left\{ 1 - \left[ A + \Delta A \, \theta(\phi-\phi_0) \right] \left( \phi - \phi_0 \right) \right\}
\end{equation}
with $A \ll 1$ so that de Sitter space is a good approximation, and $\Delta A/A \ll 1$ to ensure approximate scale invariance.
For this case
\begin{equation}\label{kf}
\ln \frac{1}{f^2} = \ln \left(\frac{V_0}{12 \pi^2 A^2}\right)
- 2 \left(\frac{\Delta A}{A}\right) \left( 1 - \frac{\xi^3}{\xi_0^3} \right) \theta(\xi_0-\xi)
+ \left(\frac{\Delta A}{A}\right)^2 \left( 1 - \frac{\xi^3}{\xi_0^3} \right)^2 \theta(\xi_0-\xi)
\end{equation}
and
\begin{equation}\label{kP}
\ln\mathcal{P} = \ln \left(\frac{V_0}{12\pi^2\,A^2} \right)
+ 2 \left(\frac{\Delta A}{A}\right) W(k\xi_0)
+ \left(\frac{\Delta A}{A}\right)^2 \left\{ 1 + X(k\xi_0) \left[ 2 \, X(k\xi_0) - \frac{3}{k^3 \xi_0^3} - \frac{3}{k \xi_0} \right] \right\}
\end{equation}
where $W(x)$ is defined in Eq.~(\ref{W}) and
\begin{equation}
X(x) = \frac{3}{2 x^3} + \frac{3}{2x} - \frac{3\cos(2x)}{2x^3} - \frac{3\sin(2x)}{x^2} + \frac{3\cos(2x)}{2x}
\end{equation}
Now, using our inverse formula Eq.~(\ref{soinv}), Eq.~(\ref{id}), and the identities
\begin{eqnarray}
\int_0^\infty \frac{dk}{k} \, m(k\xi) \left[\frac{3 \left( 1 + k^2 \xi_0^2 \right) X(k\xi_0)}{k^3 \xi_0^3} \right]
& = & 1 - \left( 1 - \frac{\xi^3}{\xi_0^3} \right)^2 \theta(\xi_0-\xi)
+ \frac{24\xi_0}{35\xi} \, \theta(\xi-\xi_0)
\nonumber \\ && \mbox{}
+ 3 \left( \frac{\xi^2}{\xi_0^2} - \frac{6\xi^4}{5\xi_0^4} + \frac{3\xi^6}{7\xi_0^6} \right) \theta(\xi_0-\xi)
\end{eqnarray}
\begin{equation}
\frac{1}{2\pi} \int_0^\infty \frac{dl}{l} \ln\left|\frac{k+l}{k-l}\right| \left[ - l\xi_0 \, W'(l\xi_0) \right]
= - X(k\xi_0)
\end{equation}
and
\begin{eqnarray}
\lefteqn{
4 \int_0^\infty \frac{dl}{l} \int_0^\infty \frac{dq}{q} \, M(l\xi,q\xi)
\left[ - l\xi_0 \, W'(l\xi_0) \right] \left[ - q\xi_0 \, W'(q\xi_0) \right]
} \nonumber \\
& = & \frac{24\xi_0}{35\xi} \, \theta(\xi-\xi_0) + 3 \left( \frac{\xi^2}{\xi_0^2} - \frac{6\xi^4}{5\xi_0^4} + \frac{3\xi^6}{7\xi_0^6} \right) \theta(\xi_0-\xi)
\end{eqnarray}
we can invert the power spectrum of Eq.~(\ref{kP}) to recover Eq.~(\ref{kf}).

\subsection*{Acknowledgements}

This work was supported in part by the Astrophysical Research Center for the Structure and
Evolution of the Cosmos funded by the Korea Science and Engineering Foundation and the Korean
Ministry of Science, the KOSEF grant KOSEF R01-2005-000-10404-0, and Brain Korea 21.

\end{document}